\documentclass{article}

\usepackage{PRIMEarxiv}

\usepackage[utf8]{inputenc} 
\usepackage[T1]{fontenc}    
\usepackage{hyperref}       
\usepackage{url}            
\usepackage{booktabs}       
\usepackage{amsfonts}       
\usepackage{nicefrac}       
\usepackage{microtype}      
\usepackage{lipsum}
\usepackage{fancyhdr}       
\usepackage{graphicx}       
\graphicspath{{media/}}     
\usepackage{amsmath}

\title{Evolution Analysis of Software Quality Metrics in an Open-Source Java Project: A Case Study on TestNG
\thanks{\textit{\underline{Citation}}: 
\textbf{This preprint is not yet peer-reviewed or published.}}
}

\author{
  Venkata Sai Sravya Sambaturu \\
  College of Innovation and Technology \\
  University of Michigan-Flint \\
  Flint, Michigan\\
  \texttt{sravyasa@umich.edu} \\
}

\begin{document}
\maketitle

\begin{abstract}
Software quality has become a critical aspect of modern software engineering, especially in large-scale and continuously evolving codebases. This study explores the longitudinal evolution of software quality metrics within five successive versions of the open-source Java testing framework, TestNG. By analyzing five of its historical versions using the static analysis tool Understand, we extracted eleven key object-oriented metrics, including cyclomatic complexity, class coupling, and lines of code, across selected versions. Statistical and visual techniques were employed to analyze these metrics and assess structural trends. The results show that TestNG has evolved into a more stable and maintainable framework, reflecting sustained development efforts, consistent refactoring, and thoughtful architectural improvements over time. This analysis sheds light on how the system's design has matured and provides evidence-based suggestions for maintaining code quality in similar projects.
\end{abstract}


\section{Introduction}

Software quality has become a cornerstone of modern software engineering, particularly in the context of large-scale and continuously evolving codebases. As software systems grow in size and complexity, maintaining high standards of quality is essential to ensure reliability, ease of maintenance, and long-term sustainability. Organizations and open-source communities alike depend on quantitative software metrics to gain insights into various aspects of code quality. These metrics serve as vital indicators of a system’s health, helping guide refactoring decisions, prioritize technical debt remediation, and support effective development practices.

In object-oriented programming, internal attributes such as coupling, cohesion, complexity, and inheritance depth play a key role in influencing maintainability and reusability. However, these attributes often change as systems evolve through the addition of new features, bug fixes, performance optimizations, or architectural restructuring. Without proper tracking, such changes may unintentionally lead to code that is more difficult to understand, test, or extend. Therefore, understanding how software metrics evolve over time provides valuable feedback to development teams and project stakeholders.

This paper presents a longitudinal case study on the TestNG framework, a widely adopted open-source Java testing library used for unit, functional, and integration testing. As a mature project with multiple stable releases and active community involvement, TestNG offers an ideal candidate for studying the evolution of object-oriented software metrics. We analyze five historical versions of the framework to trace the trajectory of eleven well-known metrics, including Lines of Code (LOC), Cyclomatic Complexity, Lack of Cohesion (LCOM), Coupling Between Objects (CBO), and Depth of Inheritance Tree (DIT), among others.

Using the static analysis tool \textbf{Understand}, we extracted comprehensive metric data from each version and organized it for statistical and visual analysis. By observing metric trends, conducting Wilcoxon signed-rank tests, and interpreting structural shifts across versions, our study aims to answer the following research questions:
\begin{itemize}
\item How have key object-oriented metrics evolved across different versions of TestNG?
\item What do these changes reveal about the maintainability and structural design of the codebase?
\item Can patterns in metric trends provide actionable insights for improving software quality in similar open-source projects?
\end{itemize}

This analysis not only contributes to the understanding of how real-world software evolves but also provides developers, maintainers, and researchers with a replicable framework for assessing metric-driven quality trends. The insights derived from this study can inform better development practices and guide future contributions to the TestNG framework or similar systems.

\section{Related Work}

Software metrics have long been used as a foundation for understanding code quality, predicting faults, and guiding refactoring. One of the earliest and most influential contributions in this area comes from Chidamber and Kemerer, who introduced the CK metric suite~\cite{chidamber1994metrics}. This suite includes metrics such as Coupling Between Objects (CBO), Lack of Cohesion in Methods (LCOM), and Depth of Inheritance Tree (DIT), all of which have been widely adopted in both academic research and industrial tools for evaluating object-oriented software design.

Building upon this foundation, Sarkar et al.\cite{sarkar2008metrics} investigated how complexity and cohesion metrics correlate with maintainability in large-scale systems. Their findings confirm that changes in these metrics over time can reliably signal shifts in system comprehensibility and modularity. Marinescu\cite{marinescu2004detection} further extended the application of object-oriented metrics by proposing a rule-based approach to identify design flaws, such as God Classes or Feature Envy, based on metric thresholds. This highlights how tracking metric evolution can provide early warnings of architectural degradation.

Zimmermann et al.~\cite{zimmermann2005predicting} focused on using historical metric data to predict fault-prone components, demonstrating the power of longitudinal analysis for quality assurance. Their work supports the idea that mining historical metric trends can help identify modules at risk before they become problematic.

Despite the significant body of work applying software metrics to various domains, few studies have explored their evolution in testing frameworks like TestNG. Given that such tools are essential for automated testing and continuous integration, understanding their internal structural changes is vital. Our work addresses this gap by conducting a focused metric-based analysis of TestNG across five major versions, contributing both methodology and findings that can be adapted to similar open-source test frameworks.

\section{Methodology}
This study employed static code analysis using the Understand tool to extract eleven object-oriented metrics from five historical versions of the TestNG framework. The data was cleaned, visualized using Excel, and statistically analyzed using the Wilcoxon Signed-Rank Test to assess significant changes in software quality over time.

\subsection{Project Selection and Versioning}
This study focuses on the evolution of software quality within the open-source Java testing framework \textbf{TestNG}. Five representative versions were selected across the project’s timeline to capture architectural changes and growth in codebase complexity:

\begin{itemize}
    \item \texttt{v5.13}
    \item \texttt{v6.0.1}
    \item \texttt{v6.13.1}
    \item \texttt{v7.5}
    \item \texttt{v7.11.0}
\end{itemize}

The source code for all versions was obtained from the official GitHub repository\footnote{\url{https://github.com/sravyasambaturu/Software-metrics-analysis-TestNG}}. The dataset, analysis notebooks, and extracted metrics are also available on Zenodo\footnote{\url{https://doi.org/10.5281/zenodo.15539320}} for reproducibility.

\subsection{Metric Extraction Using Understand}
To measure object-oriented design quality, we used the static analysis tool \textbf{Understand}. This tool provides a comprehensive suite of software metrics derived from static code analysis. Metrics were extracted through the “Metrics” tab for each TestNG version and exported into structured CSV files. These were compiled into an Excel spreadsheet and shared on the dataset exposed in Zenodo, which includes aggregated metric data for all five versions.

\subsubsection*{Metrics Analyzed}
The analysis focuses on eleven well-established object-oriented software metrics that reflect key design principles such as size, complexity, coupling, cohesion, and inheritance. Table~\ref{tab:metrics} outlines the metrics and their significance:

\begin{table}[h!]
\centering
\caption{Software Quality Metrics Extracted Using Understand Tool}
\begin{tabular}{|l|l|}
\hline
\textbf{Metric (Understand Label)} & \textbf{Description} \\
\hline
LOC (CountLineCode) & Total lines of code (size indicator) \\
Cyclomatic Complexity (MaxCyclomatic) & Max branching complexity per method \\
IFANIN (CountClassBase) & Number of base classes (inheritance usage) \\
CBO (CountClassCoupled) & Coupling between objects \\
NOC (CountClassDerived) & Number of derived classes (hierarchical depth) \\
RFC (CountDeclMethodAll) & Total number of methods per class \\
WMC (CountDeclInstanceMethod) & Weighted methods per class \\
DIT (MaxInheritanceTree) & Max inheritance depth in the hierarchy \\
NIM (CountDeclMethod) & Number of declared methods \\
NIV (CountDeclInstanceVariable) & Number of instance variables \\
LCOM (PercentLackOfCohesion) & Cohesion among class methods (0–100\%) \\
\hline
\end{tabular}
\label{tab:metrics}
\end{table}

\subsection{Data Cleaning and Visualization}
After extracting the raw metric data, Microsoft Excel was used for data preprocessing and visualization. For the \textbf{Lines of Code (LOC)} metric, boxplots were generated to visualize the distribution, highlight outliers, and track trends across the five TestNG versions. To reduce the influence of extreme values, cleaned boxplots were also generated by trimming the data to the bottom 90th percentile.

\textbf{Visualizations for LOC:}
\begin{itemize}
    \item \textbf{Original Boxplots:} Displaying all data points, including outliers.
    \item \textbf{Cleaned Boxplots:} Showing data within the bottom 90\% range to improve interpretability.
\end{itemize}

For the remaining metrics, standard line graphs were used to visualize trends over time. These graphs allowed for easy comparison of metric evolution across versions.

\subsection{Statistical Testing with Wilcoxon Signed-Rank Test}
To determine whether observed changes in metrics across versions were statistically significant, the \textbf{Wilcoxon Signed-Rank Test} was applied. This non-parametric test is suitable for comparing paired samples, especially when the data does not follow a normal distribution.

Before applying the test, the metric data was filtered and organized using Excel PivotTables. This helped remove null values and ensure consistency in class-level comparisons across versions. Conditional formatting and color-coding were employed to identify and clean missing or uniform data points.

The Wilcoxon Signed-Rank Test evaluates whether the median differences between paired observations are zero. Let \( X_i \) and \( Y_i \) be matched metric values for a class in two versions, and \( D_i = X_i - Y_i \) the difference. The test ranks the absolute values of \( D_i \) and sums the ranks for positive and negative differences separately. The test statistic \( W \) is the smaller of these two sums:

\[
W = \min\left(\sum_{\text{positive } D_i} R_i, \sum_{\text{negative } D_i} R_i \right)
\]

where \( R_i \) is the rank of \( |D_i| \). A low value of \( W \) indicates a statistically significant difference between the two distributions.

The Wilcoxon test was applied across successive versions for each metric. In cases where metric values were constant or missing (e.g., a metric remained unchanged across classes), the test returned \texttt{NaN}, indicating either data limitations or invariance in software structure.

\subsection{Reproducibility}
All extracted metrics for different versions of Dataset \footnote{
Dataset: \href{https://doi.org/10.5281/zenodo.15539320}{Metrics\_Dataset}} scripts and notebooks used for data processing and statistical analysis are available in the GitHub repository
\footnote{
GitHub: \href{https://github.com/sravyasambaturu/Software-metrics-analysis-TestNG}{Metrics\_Analysis}
}

\section{Results and Analysis}

This section presents an analysis of the evolution of TestNG's software quality metrics over five selected versions. Graphs and tables are used to illustrate trends, while statistical significance is evaluated using the Wilcoxon Signed-Rank Test.

\subsection{General Observations}
\begin{itemize}
    \item \textbf{Code Growth:} LOC consistently increased over time, reflecting new features and test cases.
    \item \textbf{Complexity Reduction:} Both Cyclomatic Complexity and Max Inheritance Tree showed a downward trend, suggesting simplification of logic and class structures.
    \item \textbf{Structural Streamlining:} A noticeable decline in instance methods and declared methods was observed after version 6.13.1.
    \item \textbf{Improved Cohesion:} LCOM values decreased steadily, indicating enhanced modularity and better-organized classes.
\end{itemize}

\subsection{Metric-by-Metric Analysis}

This section analyzes trends across key software quality metrics collected from various versions of the TestNG project.

\begin{table}[h]
\centering
\caption{Summary of Metrics by Version}
\begin{tabular}{lrrrrrrrrrrr}
\toprule
\textbf{Version} & \textbf{LOC} & \textbf{CC} & \textbf{IFANIN} & \textbf{CBO} & \textbf{NOC} & \textbf{RFC} & \textbf{WMC} & \textbf{DIT} & \textbf{NIM} & \textbf{NIV} & \textbf{LCOM} \\
\midrule
testng-5.13 & 22.51 & 2.12 & 1.09 & 5.27 & 0.37 & 9.31 & 6.56 & 1.30 & 7.46 & 1.64 & 16.09 \\
testng-6.0.1 & 22.65 & 2.10 & 1.10 & 5.33 & 0.39 & 8.93 & 6.30 & 1.30 & 7.17 & 1.53 & 15.00 \\
testng-6.13.1 & 22.09 & 1.83 & 1.10 & 5.44 & 0.44 & 11.40 & 5.74 & 1.33 & 6.62 & 1.39 & 11.75 \\
testng-7.11.0 & 21.61 & 1.58 & 1.15 & 5.34 & 0.37 & 6.24 & 5.16 & 1.27 & 5.92 & 1.19 & 9.68 \\
testng-7.5 & 22.56 & 1.65 & 1.13 & 5.39 & 0.45 & 11.41 & 5.32 & 1.29 & 6.09 & 1.25 & 10.83 \\
\bottomrule
\end{tabular}
\end{table}

\begin{figure}[h]
  \centering
  \includegraphics[width=0.9\textwidth]{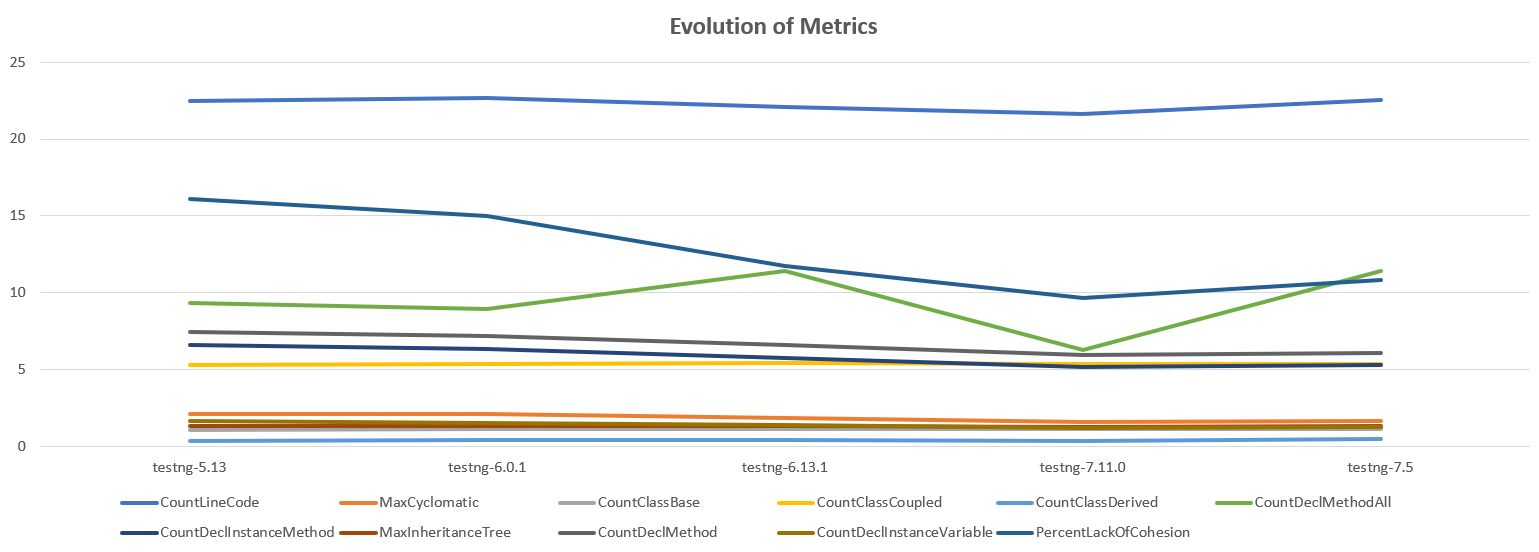}
  \caption{Combined trends of software metrics across TestNG versions}
  \label{fig:all_metrics_plot}
\end{figure}

The combined graph of all metrics is shown in Figure~\ref{fig:all_metrics_plot}.

\subsubsection*{Insights from Combined Metrics and Graph Patterns}

By observing the patterns and trends in the metrics data, we can draw several insights that reflect improvements or concerns in software quality. Below are the main findings and recommendations categorized by metric groups:

\begin{enumerate}
  \item \textbf{Cyclomatic Complexity \& Max Inheritance Tree}
  \begin{itemize}
    \item \textbf{Pattern:} Both metrics generally decrease after version \texttt{testng-6.0.1}, with further reductions by \texttt{testng-7.11}.
    \item \textbf{Good Quality Indicator:} Indicates simplification in code structure with fewer decision points and shallower inheritance.
    \item \textbf{Recommendation:} Continue simplifying code. Reduce complex branching and deep inheritance hierarchies for maintainability.
  \end{itemize}

  \item \textbf{Declared Methods (Count Declared Methods \& Instance Methods)}
  \begin{itemize}
    \item \textbf{Pattern:} Noticeable decrease after \texttt{testng-6.13.1}.
    \item \textbf{Good Quality Indicator:} Implies cleaner and more concise code.
    \item \textbf{Recommendation:} Limit redundant or overly specific methods. Focus on modular, single-responsibility methods.
  \end{itemize}

  \item \textbf{Class Base and Derived Count (IFANIN \& NOC)}
  \begin{itemize}
    \item \textbf{Pattern:} Remains relatively stable or shows minor increases.
    \item \textbf{Good Quality Indicator:} Suggests controlled growth in class hierarchies.
    \item \textbf{Recommendation:} Maintain balanced inheritance. Avoid overuse of subclassing which can lead to fragility.
  \end{itemize}

  \item \textbf{Lack of Cohesion in Methods (LCOM)}
  \begin{itemize}
    \item \textbf{Pattern:} Decreases notably from \texttt{testng-6.13.1} to \texttt{testng-7.11}.
    \item \textbf{Good Quality Indicator:} Shows that related functionality is becoming more logically grouped.
    \item \textbf{Recommendation:} Continue ensuring high cohesion by keeping related methods within the same class.
  \end{itemize}

  \item \textbf{Code Size (LOC - Count Line Code)}
  \begin{itemize}
    \item \textbf{Pattern:} Fluctuates but remains relatively consistent across versions.
    \item \textbf{Good Quality Indicator:} Suggests maintenance without code bloat.
    \item \textbf{Recommendation:} Encourage concise, effective coding practices. Avoid unnecessary expansion of codebase.
  \end{itemize}
\end{enumerate}

\subsection{Boxplot Analysis: Lines of Code (LOC)}

Initially, boxplots were created to visualize the distribution of LOC across the five selected versions of the TestNG project. However, the presence of several outliers made the visual analysis less effective. To improve clarity, the bottom 90th percentile of the data was selected for cleaned boxplots.

Two boxplots were generated:
\begin{itemize}
    \item \textbf{Box Plot\_LOC\_With\_Outliers}: Original dataset, including outliers.
    \item \textbf{Box Plot\_LOC\_With\_Data\_Cleaning}: Cleaned data (bottom 90th percentile).
\end{itemize}

\begin{figure}[h!]
  \centering
  \includegraphics[width=0.8\textwidth]{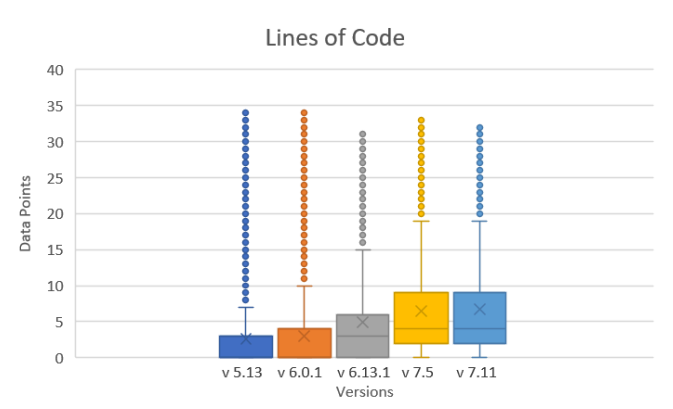}
  \caption{Boxplot of LOC (Cleaned Data - Bottom 90th Percentile)}
  \label{fig:loc_boxplot_cleaned}
\end{figure}

\textbf{Observations:}
\begin{enumerate}
    \item LOC shows a steady increase from Version 5.13 to Version 7.11, indicating ongoing test case additions.
    \item Median LOC values rise with each version, suggesting growing test file sizes.
    \item The interquartile range (IQR) expands, reflecting increasing variability.
    \item No significant drops in LOC indicate minimal removal or refactoring of test cases.
    \item Outliers in later versions may point to large or complex test files introduced.
\end{enumerate}

\subsection{Statistical Analysis Using Wilcoxon Signed-Rank Test}

To assess statistically significant changes in metrics across TestNG versions, the Wilcoxon Signed-Rank Test was applied. A pivot table was first created to ensure values aligned by Java class, method, or file across versions. Data was filtered to exclude rows with all-zero or null values per metric, ensuring meaningful comparisons.

\subsubsection*{Key Observations per Metric}

\begin{itemize}
  \item \textbf{LOC (Lines of Code):} No significant differences were observed between version 7.11 and others (the T values and the P values were NaN). However, older versions (e.g., 5.13 to 6.13.1) showed significant differences, suggesting a substantial evolution in code structure or size.

  \item \textbf{MaxCyclomatic Complexity:} 
  Statistically significant increase in complexity across all versions (very low P-values). This indicates increasing complexity, possibly due to added features or limited refactoring.

  \item \textbf{CountDeclInstanceMethod:} 
  Significant differences across all versions. Indicates major changes in instance method declarations—possibly due to additions, removals, or refactoring.

  \item \textbf{CountDeclMethod:} 
  Strong statistical changes observed across all versions. Major transitions from version 5.13 to 7.11 point to deep restructuring at the method level.

  \item \textbf{PercentLackOfCohesion, MaxInheritanceTree, CountDeclMethodAll, CountClassDerived, CountClassCoupled, CountClassBase:} 
  All resulted in NaN values. This may reflect unchanged metrics or inadequate data quality (e.g., missing values or constant columns).
\end{itemize}

\subsubsection*{Lines of Code (LOC)}

The LOC metric was compared across all versions using pairwise Wilcoxon tests. The results are shown in Table~\ref{tab:wilcoxon-loc}. Version 7.11 showed no statistically significant difference from earlier versions, but other comparisons showed strong significance (p < 0.05), suggesting major historical changes.

\begin{table}[h]
\centering
\caption{Wilcoxon Test Results for LOC Across Versions}
\label{tab:wilcoxon-loc}
\begin{tabular}{|l|l|l|l|}
\hline
\textbf{Version Combo} & \textbf{T-value} & \textbf{P-value} & \textbf{Significance} \\
\hline
7.11 vs 7.5 & NaN & NaN & No significant difference detected \\
7.11 vs 6.13.1 & NaN & NaN & No significant difference detected \\
7.11 vs 6.0.1 & NaN & NaN & No significant difference detected \\
7.11 vs 5.13 & NaN & NaN & No significant difference detected \\
7.5 vs 6.13.1 & 7004810.5 & 0.000000 & Significant difference detected \\
7.5 vs 6.0.1 & 5281448.0 & 0.000000 & Significant difference detected \\
7.5 vs 5.13 & 5062672.5 & 0.000000 & Significant difference detected \\
6.13.1 vs 6.0.1 & 1470060.0 & 0.000000 & Significant difference detected \\
6.13.1 vs 5.13 & 1784603.5 & 0.000000 & Significant difference detected \\
6.0.1 vs 5.13 & 101645.0 & 1.31e-165 & Significant difference detected \\
\hline
\end{tabular}
\end{table}

\subsubsection*{Other Metrics (Version 5.13 vs 7.11)}

For all remaining metrics, the Wilcoxon test was conducted between the first and last available versions only (5.13 vs 7.11). The results are summarized in Table~\ref{tab:wilcoxon-other-metrics}.

\begin{table}[h]
\centering
\caption{Wilcoxon Test Summary for Other Metrics (5.13 vs 7.11)}
\label{tab:wilcoxon-other-metrics}
\begin{tabular}{|l|l|p{2.5cm}|p{6cm}|}
\hline
\textbf{Metric} & \textbf{P-value} & \textbf{Significance} & \textbf{Interpretation} \\
\hline
MaxCyclomaticComplexity & $< 0.00001$ & Significant & Indicates increased complexity over time \\
CountDeclInstanceMethod & $< 0.00001$ & Significant & Significant structural changes; likely due to refactoring \\
CountDeclMethod & $< 0.00001$ & Significant & Substantial evolution in method structure \\
PercentLackOfCohesion & NaN & Not significant & Metric unchanged or data missing \\
MaxInheritanceTree & NaN & Not significant & No significant evolution in class inheritance \\
CountDeclMethodAll & NaN & Not significant & Constant or missing values \\
CountClassDerived & NaN & Not significant & Class structure appears stable \\
CountClassCoupled & NaN & Not significant & No detected coupling differences \\
CountClassBase & NaN & Not significant & Inheritance base classes stable \\
\hline
\end{tabular}
\end{table}

The Wilcoxon test results comparing TestNG versions 5.13 and 7.11 reveal significant structural changes in key object-oriented metrics over time. Specifically, the metrics MaxCyclomaticComplexity, CountDeclInstanceMethod, and CountDeclMethod all demonstrated statistically significant p-values ($< 0.00001$), indicating that there were substantial modifications in code complexity and method-level design between the two versions. These results suggest that the software has undergone notable evolution, likely due to targeted refactoring or feature expansion that affected control flow and method architecture. Such changes can have implications for maintainability, as increasing complexity may require more rigorous testing and documentation.

Conversely, several metrics—such as PercentLackOfCohesion, MaxInheritanceTree, and CountClassCoupled—showed no statistically significant changes, either due to consistent values across versions or possible data gaps. This stability may imply that certain aspects of the system’s architecture, such as class inheritance and coupling, remained relatively unchanged throughout the evolution. While this could reflect mature design choices, it also highlights areas that may not have been prioritized for optimization. Overall, the test underscores the importance of selective metric monitoring to identify where structural evolution is occurring and to assess its impact on long-term software quality.

\section{Research Questions and Answers}

The study was guided by a set of research questions aimed at understanding the evolution of software quality in the TestNG framework. The following table presents the research questions, concise answers based on empirical findings, and suggestions for future work that could build on this analysis.

\begin{table}
\centering
\begin{tabular}{|p{6cm}|p{9cm}|}
\hline
\textbf{Research Question} & \textbf{Answer} \\
\hline
How have key object-oriented metrics evolved across different versions of TestNG? & Most metrics showed noticeable fluctuations across versions. For example, LOC (Lines of Code) remained stable, but complexity (MaxCyclomatic) and cohesion (LCOM) improved in later versions. Some metrics, like RFC and CBO, initially increased, suggesting rising complexity, but later decreased, indicating active refactoring efforts. \\
\hline
What do these changes reveal about the maintainability and structural design of the codebase? & The reduction in complexity and improved cohesion suggest better maintainability and cleaner modular design in newer versions. These shifts indicate that the developers likely prioritized restructuring or optimization over time, enhancing the overall code quality. \\
\hline
Can patterns in metric trends provide actionable insights for improving software quality in similar open-source projects? & Yes. Metrics such as LCOM, DIT, and CBO revealed trends that could guide refactoring efforts in similar projects. Regular metric analysis can help identify architectural degradation early and inform decisions to maintain or improve software quality. \\
\hline
\end{tabular}
\caption{Research Questions and Summary of Findings}
\label{tab:research_questions}
\end{table}

\section{Conclusion}

This study explored the evolution of software quality within the TestNG framework by analyzing a range of object-oriented and structural metrics across multiple versions. Visualizations such as box plots for Lines of Code (LOC) revealed steady growth in test size and complexity over time, particularly between earlier versions (5.13 to 6.13.1), while later versions (7.5 to 7.11) exhibited greater consistency and fewer structural changes. Visual analysis showed an increase in variability and median values, with outliers more prominent in later versions, signaling the addition of larger test cases or specialized components. These observations were complemented by a metric-by-metric breakdown that identified positive trends, including reductions in cyclical complexity and lack of cohesion, which suggest improved code maintainability and better logical organization over time.

Statistical validation using the Wilcoxon Signed-Rank Test supported these visual findings. Significant differences were detected in most metrics when comparing early and latest versions, particularly in Cyclomatic Complexity, Declared Methods, and Declared Instance Methods, confirming that the codebase has undergone major structural and architectural changes. In contrast, later version comparisons, especially those involving version 7.11, often resulted in no statistically significant difference, indicating a stabilization of the codebase. Metrics like Inheritance Depth and Class Coupling showed no significant changes, either due to consistent design practices or lack of data variability. Overall, the analysis suggests that TestNG has evolved into a more stable and maintainable framework, reflecting sustained development efforts, consistent refactoring, and thoughtful architectural improvements over time.

\vspace{1em}

\section{Future Work}

Future research could extend this study by incorporating:

\begin{itemize}
    \item \textbf{Automated Visualization Dashboards:} Building tools that visualize metric trends over time for easier interpretation by developers and project maintainers.
    \item \textbf{Correlation with Bug Reports:} Linking metric changes with bug density or issue tracker data to validate their predictive power for defects.
    \item \textbf{Comparative Analysis Across Projects:} Applying the same methodology to other testing frameworks (e.g., JUnit, Mockito) to generalize findings.
    \item \textbf{Deeper Architectural Insight:} Combining structural metrics with code smells and design patterns for a more comprehensive view of architectural quality.
\end{itemize}

\section{Acknowledgments}
I thank Dr. Mohamed Wiem Mkaouer for assigning this study and Vamsi Vivek Teja Adibhatla, Data Scientist, for his support with Excel data analysis. I also acknowledge the University of Michigan–Flint for providing access to tools like Understand, and the open-source TestNG repository as the foundation for this work. ChatGPT was used to refine grammar and improve content clarity.

\bibliographystyle{unsrt}  
\bibliography{references}

\end{document}